\documentclass[conference]{IEEEtran}
\IEEEoverridecommandlockouts
\usepackage{cite}
\usepackage{amsmath,amssymb,amsfonts}
\usepackage{graphicx}
\usepackage{textcomp}
\usepackage{xcolor}
\usepackage{empheq}
\usepackage{algorithmicx}
\usepackage{algorithm,algpseudocode}
\usepackage{algorithmicx}
\usepackage{setspace}
\usepackage{subcaption}

\usepackage{float}
\usepackage{color}
\usepackage{url}
\usepackage{dblfloatfix} 
\usepackage{graphicx}
\usepackage{verbatim} 

\renewcommand{\baselinestretch}{0.96}

\usepackage{multirow}
\usepackage[short]{optidef}
\usepackage{balance}
\usepackage{hyperref}
\hypersetup{
    colorlinks=true,
    linkcolor=blue,
    filecolor=magenta,      
    urlcolor=cyan,
}

\def\BibTeX{{\rm B\kern-.05em{\sc i\kern-.025em b}\kern-.08em
    T\kern-.1667em\lower.7ex\hbox{E}\kern-.125emX}}
\begin{document}

\title{RF Signal Source Search and Localization Using an Autonomous UAV with Predefined Waypoints 
\thanks{This research is supported in part by the NSF award CNS-1939334.}
}

\author{\IEEEauthorblockN{Hyeokjun Kwon and Ismail Guvenc
}
\IEEEauthorblockA{Department of Electrical and Computer Engineering,
North Carolina State University, Raleigh, NC}
\IEEEauthorblockA{\{khyeokj,iguvenc\}@ncsu.edu}
}

\maketitle

\begin{abstract}
Localization of a radio frequency (RF) signal source has various use cases, ranging from search and rescue, identification and deactivation of jammers, and tracking hostile activity near borders or on the battlefield. The use of unmanned aerial vehicles (UAVs)  for signal source search and localization (SSSL) can have significant advantages when compared to terrestrial-based approaches, due to the ease of capturing RF signals at higher altitudes and the autonomous 3D navigation capabilities of UAVs. However, the limited flight duration of UAVs due to battery constraints, as well as limited computational resources on board of lightweight UAVs introduce challenges for SSSL. In this paper, we study various SSSL techniques using a UAV with predefined waypoints. 

A linear least square (LLS) based localization scheme is considered with enhanced reference selection due to its relatively lower computational complexity. Five different LLS localization algorithms are proposed and studied for selecting anchor positions to be used for localization as the UAV navigates through an area. The performance of each algorithm is measured in two ways: 1) real-time positioning accuracy during the ongoing UAV flight, and 2) long-term accuracy measured at the end of the UAV flight. We compare and analyze the performance of the proposed approaches using computer simulations in terms of accuracy, UAV flight distance, and reliability.
\end{abstract}


\section{Introduction}
Unmanned aerial vehicles (UAVs) can play a crucial role in location-based applications and remote sensing networks~\cite{leo}. Especially, using UAVs for signal source searching and localization (SSSL) has significant advantages over terrestrial-based SSSL approaches due to the ease of capturing RF signals at higher altitudes and the autonomous 3D navigation capabilities of UAVs. For example, UAVs can be utilized in military surveillance and border patrol missions that require high accuracy of signal source positioning~\cite{military}, or for localizing intentional or non-intentional jammers~\cite{perkins2015antenna}. An approximate but fast SSSL approach using UAVs can save lives during a disaster or emergency situations~\cite{disaster}. 

A wide range of studies that use UAVs for localization are available in the literature~\cite{military, perkins2015antenna, disaster, kalman, hawk, piecewise, realtime}. Finding the optimal UAV trajectory for a given area or figuring out the better formation of UAVs for localization have been studied in the past. For example, in~\cite{survel}, various UAV search patterns and their traits are summarized. Trajectory design for localizing a source signal with an unmanned mini-helicopter is investigated based on a Moore space-filling curve algorithm in~\cite{hawk}. In another recent study, a group of small UAVs performs radio frequency positioning with flexible and practical formations in~\cite{realtime}.

Most of the existing studies focus on improving the accuracy of localization. Among them, statistical approaches are extensively explored. In~\cite{ekf},  localization and tracking algorithms based on an extended Kalman filter are analyzed. Parallel with this, particle filter-based localization algorithms are investigated and compared with an extended Kalman filter algorithm in~\cite{kalman}. Recently, machine learning (ML) based localization algorithms have been getting attention for UAV-based SSSL. In~\cite{qlearn}, a Q-Learning-based positioning algorithm is suggested to localize illegal radio stations. Similar to this, source signal-based Q-learning scheme for UAV navigation is studied in~\cite{rssqlearn}. However, the aforementioned algorithms may require a high computational complexity, which may not be feasible for lightweight UAVs with limited batter and computational capacities. Moreover, existing ML-based approaches typically require prior training of the ML technique in the same environment which may not always be possible. 
In this sense, the computation complexity of various localization algorithms is studied in a broader sense in~\cite{complex}. Furthermore, efforts to reduce the computational complexity of UAV based localization are conducted by converting the nonlinear and non-convex problems into linear and convex problems in~\cite{piecewise}. 

While UAV-based SSSL techniques have been investigated in the literature, there are still various research challenges to be addressed. While some SSSL use cases may prioritize accuracy at the cost of longer search time, some other use cases may require a rough location estimate but need it quickly and at low computational complexity. In this context, the main contributions of this work can be listed as follows:
\begin{itemize}
\item Five approaches are proposed for anchor point selection over a UAV's trajectory to carry out linear least squares (LLS)-based real-time target localization; 
\item Accuracy, latency, reliability, and complexity of the five LLS-based localization techniques are compared using computer simulations;  
\item Two methods for selecting the linearization reference point are compared for LLS: i) static reference selection (SRL) that uses the first point of the UAV trajectory as the reference point, and ii) dynamic reference selection (DRL) that uses the reference point dynamically.  
\end{itemize}

The remainder of this paper is organized as follows. In Section~\ref{Sec:SysModel}, the system models of UAV, the channel model, and the theoretical base of the LLS algorithm are given. In Section~\ref{Sec:PropScheme}, five SSSL schemes are described. Simulation settings and numerical results are demonstrated in Section~\ref{Sec:NumResult}. Lastly, conclusions are given in Section~\ref{Sec:Conclusion}.

\section{System Model}\label{Sec:SysModel}

\begin{figure}[t!]
\vspace{-1mm}
\centering
\includegraphics[trim=0.2cm 0.1cm 0.2cm 0.3cm, clip,width=7.8cm]{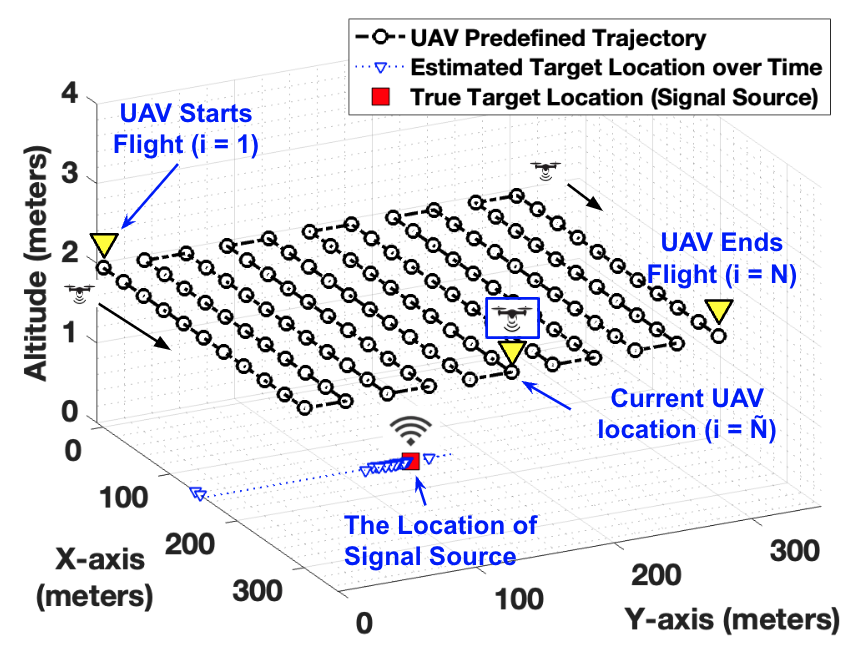}
\caption{SSSL for ground target localization using an autonomous UAV with predefined waypoints.}
\label{fig:SSSLscheme}
\vspace{-5mm}
\end{figure}

In this paper, we consider the problem of RF SSSL of a target node on the ground using a single autonomous UAV as illustrated in Fig.~\ref{fig:SSSLscheme}. In particular, a UAV flies autonomously on a path with predefined waypoints and its goal is to localize the target node as quickly and as accurately as possible. To do that, the UAV has to select anchor points for continuously estimating the target's location as it moves through its trajectory. As more observations are obtained, localization accuracy is expected to improve. However, using all reference points over the UAV's trajectory does not necessarily provide the best accuracy, as will be shown using computer simulations.  

The propagation channel model for UAV network is typically assumed to have a line of sight (LOS) with the signal source, which we also assume in this paper for simplicity. For an LOS path, accurate distance estimation is possible because path loss mainly depends on the  distance between the UAV and the transmitter~\cite{indoor}. 
Considering LOS and open space, free-space path loss (FSPL) model is used to calculate the estimated distance between the target node and the UAV, given by~\cite{uavcom}:
\begin{equation}\label{eq:1}
{\rm PL}(d) = {\rm PL}_0 + 10\alpha\log_{10} \left( \frac{d}{d_0} \right)+X_\sigma,
\end{equation}
where ${\rm PL}_0$ is the path loss at the reference distance $d_0$, $\alpha$ denotes the path loss exponent in the free space that is set to be 2 in this work, $X_\sigma$ is a Gaussian random variable with mean zero and standard deviation $\sigma$ to capture shadowing effects, and $d$ is the distance between the ground target and the UAV. Using~\eqref{eq:1}, the distance $\hat{d}_i$ with the ground target node can be estimated at the $i$th UAV flight location and it can then be expressed as follows~\cite{funda}:
\begin{equation}\label{eq:2}
\hat{d}_i = f_{i}(x,y,z) + n_{i},\;\;\; i = 1,...,\tilde{N},\;\;\; \tilde{N} = 1,...,N~,
\end{equation}
where $n_{i}$ is the white noise at the $i$th UAV location, $\tilde{N}$ is the index for the current location of the UAV, $N$ is the index of the last location at the end of the UAV's trajectory, and $f_{i}(x,y,z)$ is the real distance between the target $(x,y,z)$ and the $i$th UAV location $(x_{i},y_{i},z_{i})$. This can be described as~\cite{funda} 
\begin{equation}\label{eq:3}
f_{i}(x,y,z) = \sqrt{(x-x_{i})^2 + (y-y_{i})^2 + (z-z_{i})^2}.
\vspace{+0.2cm }
\end{equation}

With the estimated distance $\hat{d}_i$ obtained at multiple UAV reference points, the least squares (LS) algorithm can be effectively used for both linear and nonlinear parameter estimation. However, due to the high computational complexity of the nonlinear least square algorithm~\cite{funda}, we consider an LLS-based localization technique in this paper.
In the LLS algorithm, a set of linear equations are obtained based on simple mathematical combinations of equations \eqref{eq:2} and \eqref{eq:3}. 
In particular, \eqref{eq:2} and \eqref{eq:3} can be expressed (neglecting noise terms) as follows for $i=1,...,\tilde{N}$~\cite{3d}
\begin{equation}\label{eq:4}
d_{i}^2 = \left( x - x_{i} \right)^2 + \left( y - y_{i} \right)^2 + \left( z - z_{i} \right)^2~.
\end{equation}
Localization of the ground target requires distance estimates to the target measured at a minimum of three different UAV anchor positions. 
We define the index set $\boldsymbol{\upsilon}$ of size $3\leq S\leq \tilde{N}$ that captures the unique UAV indices (anchor positions) to be used in the localization process. Details of how to form $\boldsymbol{\upsilon}$ will be discussed further in the next section.  

For LLS-based localization, we have to set the $r$th index, $r\in\boldsymbol{\upsilon}$, as a reference index for localization. Two methods are suggested for selecting the linearization reference. With the SRL mode, the first index of $\boldsymbol{\upsilon}$ is selected as the reference. On the other hand, in DRL mode, indices in the $\boldsymbol{\upsilon}$ are sorted in ascending order, and the index with the smallest distance (between the target and the UAV) is adopted as the reference. This reference selection scheme is used for all the suggested LLS schemes in Section~\ref{Sec:PropScheme}.

After selecting the reference for linearization, the reference equation (corresponding to $\hat{d}_r$ in~\eqref{eq:2}) is subtracted from all the other equations corresponding to the set $\boldsymbol{\upsilon}$, which can be written in matrix notation as follows for $k=1,\cdots,{S}$,~and~$k\neq r$~\cite{funda,3d}   
\begin{equation}\label{eq:5}
\boldsymbol{A}_{\boldsymbol{\upsilon}}\boldsymbol{l} = \boldsymbol{b}_{\boldsymbol{\upsilon}}~,
\end{equation}
where $\boldsymbol{l}=[\hat{x},\:\hat{y},\:\hat{z}]^T$ is the estimated target location,
\begin{equation}\label{eq:6}
\boldsymbol{A}_{\boldsymbol{\upsilon}} = 2
\begin{bmatrix}
x_{{\boldsymbol{\upsilon}(1)}}-x_{r} & y_{{\boldsymbol{\upsilon}(1)}}-y_{r} & z_{{\boldsymbol{\upsilon}(1)}}-z_{r} \\
\; & \vdots & \; \\
x_{{\boldsymbol{\upsilon}(k)}}-x_{r} & y_{{\boldsymbol{\upsilon}(k)}}-y_{r} & z_{{\boldsymbol{\upsilon}(k)}}-z_{r} \\
\; & \vdots & \; \\
x_{{\boldsymbol{\upsilon}(S)}}-x_{r} & y_{{\boldsymbol{\upsilon}(S)}}-y_{r} & z_{{\boldsymbol{\upsilon}(S)}}-z_{r} 
\end{bmatrix}~,
\end{equation}
and
\begin{equation}\label{eq:7}
\boldsymbol{b}_{\boldsymbol{\upsilon}} = 
\begin{bmatrix}
d_{r}^2 - d_{{\boldsymbol{\upsilon}(1)}}^2 + x_{{\boldsymbol{\upsilon}(1)}}^2 + y_{{\boldsymbol{\upsilon}(1)}}^2 + z_{{\boldsymbol{\upsilon}(1)}}^2 - \lambda \\
\vdots \\
d_{r}^2 - d_{{\boldsymbol{\upsilon}(k)}}^2 + x_{{\boldsymbol{\upsilon}(k)}}^2 + y_{{\boldsymbol{\upsilon}(k)}}^2 + z_{{\boldsymbol{\upsilon}(k)}}^2 - \lambda \\
\vdots \\
d_{r}^2 - d_{{\boldsymbol{\upsilon}(S)}}^2 + x_{{\boldsymbol{\upsilon}(S)}}^2 + y_{{\boldsymbol{\upsilon}(S)}}^2 + z_{{\boldsymbol{\upsilon}(S)}}^2 - \lambda
\end{bmatrix}~,
\vspace{+1mm}
\end{equation}
where $\lambda$ is ($x_{r}^2 + y_{r}^2 + z_{r}^2$) and $\boldsymbol{\upsilon}(k)$ is $k$th index of  $\boldsymbol{\upsilon}$. The estimated target location $\boldsymbol{l}$ is given using LLS as \cite{funda, 3d}
\begin{equation}\label{eq:8}
\boldsymbol{l}= (\boldsymbol{A}_{\boldsymbol{\upsilon}}^T\boldsymbol{A}_{\boldsymbol{\upsilon}})^{-1}\boldsymbol{A}_{\boldsymbol{\upsilon}}^T\boldsymbol{b}_{\boldsymbol{\upsilon}}~.
\end{equation}

\begin{table}[t]\label{t1}
\caption{Summary of parameters for generic LLS algorithm.}
\begin{center}
\scalebox{0.75}{
\begin{tabular}{|c|c|}
\hline
Parameters        & Stands for                                       \\ \hline
$i$               & UAV flight index                            \\ \hline
$\rm{N}$          & The index at the end of the UAV trajectory        \\ \hline
$\tilde{N}$       & The UAV index at the current location    \\ \hline
$\boldsymbol{\upsilon}$        & a set of UAV indices used for LLS algorithm      \\ \hline
$S$          & Size of index array $\boldsymbol{\upsilon}$ \\ \hline
$k$          & Elements index in $\boldsymbol{\upsilon}$    \\ \hline
\end{tabular}}
\end{center}
\vspace{-5mm}
\end{table}

\section{Proposed SSSL Schemes}\label{Sec:PropScheme}

Based on the system model in Section~\ref{Sec:SysModel}, in this section we propose and evaluate five different approaches for dynamically forming and updating the index set $\boldsymbol{\upsilon}$ as the UAV moves along its trajectory. In particular, we study five different approaches: 1) Consecutive three-point (CON); 2) Cumulative all-points (CUM); 3) First-mid-last points (FML); 4) Convex hull modified (CHLM); and 5) Closest three-points (CLS). Each algorithm considers  different approaches for obtaining anchors and the reference UAV positions to populate the matrix $\boldsymbol{A}_{\boldsymbol{\upsilon}}$ in~\eqref{eq:6}. 
After the anchors are selected for the current location $\tilde{N}$ of the UAV, LLS based localization is carried out with SRL or DRL as described in the previous section. 

\subsection{LLS-CON}
The main idea of the LLS-CON is to use the minimum relative residual, where the relative residual is defined as: 
\begin{equation}\label{eq:10}
R_{\tilde{N}} = \frac{\left\|\boldsymbol{b}_{\boldsymbol{\upsilon}} - \boldsymbol{A}_{\boldsymbol{\upsilon}}{\boldsymbol{l}}\right\|}{\left\| \boldsymbol{b}_{\boldsymbol{\upsilon}} \right\|}~,
\end{equation}
where $\left\|.\right\|$ returns the $L_{1}$ norm of its argument, and $\boldsymbol{l}$ is the estimated target location with LLS localization based on the UAV index array $\boldsymbol{\upsilon}$. In this algorithm, three consecutive points are used as anchor locations in $\boldsymbol{\upsilon}$ for SSSL. In particular, the index array $\boldsymbol{\upsilon}$ is defined as follows:
\begin{equation}
\boldsymbol{\upsilon}\:=
\begin{cases}
\big\{ \tilde{N}, \: \tilde{N}-1, \:\tilde{N}-2 \big\}, & \;\; {\rm if} \; R_{\tilde{N}}\: \leq \: R_{\tilde{N}-1} \\
\big\{ \tilde{N}-1, \: \tilde{N}-2, \:\tilde{N}-3 \big\}, & \;\; {\rm otherwise}
\end{cases}~.\nonumber 
\end{equation}
\vspace{+0.1mm}
Two modified versions of LLS-CON  are also considered that leave larger gaps between consecutive UAV locations. In particular, $\boldsymbol{\upsilon}$ for LLS-CON-I is defined as
\begin{equation}
\boldsymbol{\upsilon}\:=
\begin{cases}
\big\{{ \tilde{N}, \: \tilde{N}-2, \:\tilde{N}-4 \big\}}, & \;\; {\rm if} \; R_{\tilde{N}}\: \leq \: R_{\tilde{N}-1} \\
\big\{{ \tilde{N}-1, \: \tilde{N}-3, \:\tilde{N}-5 \big\}}, & \;\; {\rm otherwise}
\end{cases}~,\nonumber
\end{equation}
while for LLS-CON-II, $\boldsymbol{\upsilon}$ is defined as
\begin{equation}
\boldsymbol{\upsilon}\:=
\begin{cases}
\big\{ \tilde{N}, \: \tilde{N}-3, \:\tilde{N}-6 \big\}, & \;\; {\rm if} \; R_{\tilde{N}}\: \leq \: R_{\tilde{N}-1} \\
\big\{ \tilde{N}-1, \: \tilde{N}-4, \:\tilde{N}-7 \big\}, & \;\; {\rm otherwise}
\end{cases}~.\nonumber
\end{equation}
Fig.~\ref{fig:al1a} summarizes how LLS-CON and its variations select $\boldsymbol{\upsilon}$ on the UAV's trajectory. 

\subsection{LLS-CUM}
\begin{figure}[!t]
\centering
\begin{subfigure}{0.78\columnwidth}
\centering
\includegraphics[width=\textwidth]{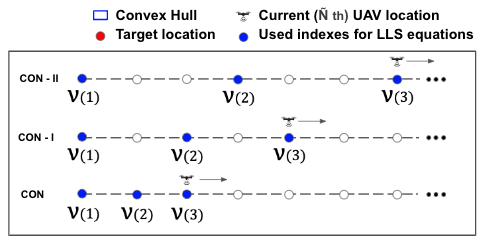}
\caption{LLS-CON}
\label{fig:al1a}
\end{subfigure}
\begin{subfigure}{0.78\columnwidth}
\centering
\includegraphics[width=\textwidth]{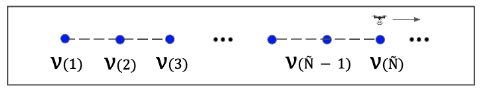}
\caption{LLS-CUM}
\label{fig:al1b}
\end{subfigure}
\begin{subfigure}{0.78\columnwidth}
\centering
\includegraphics[width=\textwidth]{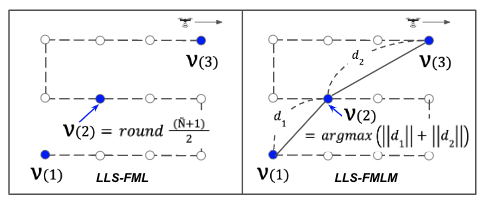}
\caption{LLS-FML}\label{fig:al1c}
\end{subfigure} 
\begin{subfigure}{0.78\columnwidth}
\centering
\includegraphics[width=\textwidth]{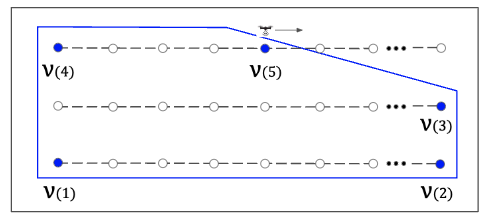}
\caption{LLS-CHLM}\label{fig:al1d}
\end{subfigure}
\begin{subfigure}{0.78\columnwidth}
\centering
\includegraphics[width=\textwidth]{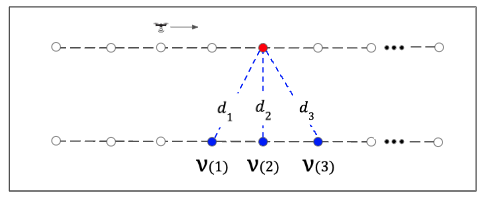}
\caption{LLS-CLS}\label{fig:al1e}
\end{subfigure}
\caption{Five different SSSL algorithms considered in this paper.}
\label{fig:algorithm}
\vspace{-5mm}
\end{figure}

In LLS-CUM, all the indices of the UAV trajectory are used, rather than a specific number of indices. More specifically, the index set for calculating the location estimate is given by $\boldsymbol{\upsilon}\:=
\big\{ 1, \:..., \:\tilde{N}\big\}$.
The size of the matrices $\boldsymbol{A}_{\boldsymbol{\upsilon}}$ and $\boldsymbol{b}_{\boldsymbol{\upsilon}}$ in \eqref{eq:8} at index $\tilde{N}$ increases to $(\tilde{N}-1) \times\: 3$ and $(\tilde{N}-1) \times\: 1$, respectively. The increasing size of the parameter matrices gradually increases the computational complexity as the UAV moves along its trajectory. Fig.~\ref{fig:al1b} highlights how LLS-CUM forms the set $\boldsymbol{\upsilon}$ over a UAV's trajectory. 

\subsection{LLS-FML}
The main goal of the LLS-FML is to maximize the gap between each index in the set $\boldsymbol{\upsilon}$. In this algorithm, three indices are used for localization, and the index set is defined as:
\begin{equation}
\boldsymbol{\upsilon}~=
\left\{ 1, \: {\rm round}\left(\frac{\tilde{N}+1}{2}\right), \:\tilde{N} \right\}~, \;\; \tilde{N}\geq 3~.\nonumber
\end{equation}
We also study a modified version of LLS-FML, which we refer as LLS-FMLM, which selects the reference locations as 
\begin{equation}
\boldsymbol{\upsilon}~=
\big\{ 1, \: m, \:\tilde{N} \big\}~, \;\; \tilde{N}\geq 3~,\nonumber
\end{equation}
where the index $m$ is calculated as
\begin{equation}\label{eq:12-a}
m = \arg \underset{m}{\operatorname{max}}~\big(d_{1m} \:+\:  d_{m\tilde{N}} \big)~,
\end{equation}
where $d_{1m} = \left\|(x_{1}, y_{1}, z_{1}) - (x_{\rm m}, y_{\rm m}, z_{\rm m})\right\|$ and $d_{m\tilde{N}} = \left\|(x_{\tilde{N}}, y_{\tilde{N}}, z_{\tilde{N}}) - (x_{\rm m}, y_{\rm m}, z_{\rm m})\right\|$.
The aim of LLS-FMLM is to maximize the physical total distance separation between the second UAV anchor location and the other two UAV anchor locations. For both LLS-FML and LLS-FMLM, the index array $\boldsymbol{\upsilon}$ results in various geometric shapes for the UAV anchor locations, and the localization performance varies based on those geometries. Fig.~\ref{fig:al1c} illustrates the operation of LLS-FML and LLS-FMLM.

\subsection{LLS-CHLM}
LLS-CHLM is a geometric approach that exploits the convex hull of the UAV's trajectory. At least three and a maximum of five points of UAV flight are used for generating the set $\boldsymbol{\upsilon}$ in this model. When every UAV trajectory is on the same line, a convex is not formed, and for this specific case, the FML algorithm is used. When convex shapes are formed on the UAV's trajectory, a standard algorithm for the convex hull~\cite{comgeo} is used to get the boundary location indices that contain all other trajectory points. In this scheme, only the corner points of a convex hull are used to reduce the computational complexity, as shown in Fig.~\ref{fig:al1d}. 

\subsection{LLS-CLS}
The LLS-CLS takes advantage of the lower path loss at the  UAV anchor points that are \emph{closer} to the ground target to be localized.  
In this model, the three closest indices to the target location are used to form LLS equations. For CLS, the index array $\boldsymbol{\upsilon}$ is formed as
\begin{equation}
\boldsymbol{\upsilon}\:=
\big\{ \boldsymbol{\upsilon}_{\phi}(1), \: \boldsymbol{\upsilon}_{\phi}(2), \: \boldsymbol{\upsilon}_{\phi}(3) \big\}~,\nonumber
\end{equation}
where $\boldsymbol{\upsilon}_{\phi}$ is a reordered version of all candidate UAV locations $\{1,...,\tilde{N}\}$, sorted by ascending measurement distances to the target node. In other words, LLS-CLS always uses the three UAV locations that have the closest measurement distance to the target. Note that once the UAV is in the vicinity of the target, the closest three locations are expected to provide a very small localization error. However, it may take a long time for the UAV to fly close to the target depending on the target's location. Moreover, antenna radiation patterns, shadowing, and small-scale fading may negatively affect the accuracy of distance estimates, and hence the accuracy of LLS-CLS, which are not explicitly studied in this paper.  Fig.~\ref{fig:al1e} describes the operational principle of LLS-CLS.

\subsection{Performance Analysis Criteria}

Performances of each of the five methods described earlier in this section are evaluated in multiple different ways. First, we evaluate the evolution of the root mean square error (RMSE) of real-time localization as the UAV moves along its trajectory. This provides how fast the localization accuracy gets improved for each approach. Second, we evaluate the RMSE at the endpoint of the UAV trajectory, which we refer to in this paper as the \emph{long-term} RMSE, and average the performance over different target locations. We also study the cumulative distribution function (CDF) of the long-term RMSE for all scenarios. 

\begin{figure}[t!]
\centering
\includegraphics[trim=0.2cm 0.2cm 0.2cm 0.2cm, clip,width=6cm]
{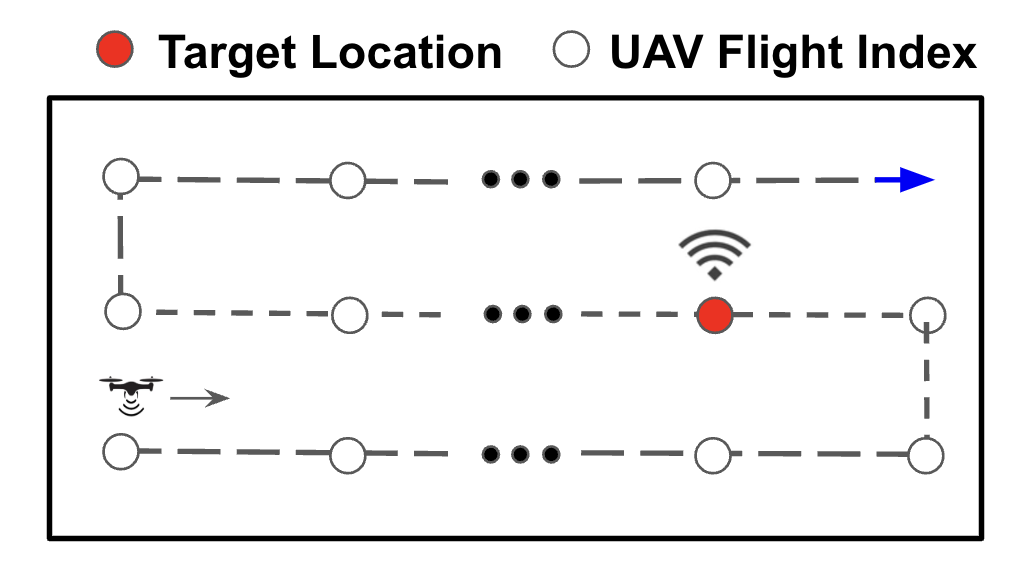}
\caption{Parallel track pattern for SSSL with a UAV.}
\label{fig:trajec}
\vspace{-4mm}
\end{figure}

The comparison of the localization performance is based on three criteria. First, localization accuracy is considered, as it is critical for use cases that prioritize high precision over search time. The second criterion is the UAV flight distance as it will directly affect the search time. To measure the flight distance, we define a minimally acceptable localization RMSE threshold ($\theta$) as $20$~m. A ground target is assumed to be localized if the localization error is lower than, and does not again become higher than, $\theta$. We define the flight distance when this accuracy is achieved as the required minimum flight distance. 
The third criterion is reliability. If the RMSE of an algorithm fluctuates as the UAV moves through its trajectory, the algorithm is considered more unreliable even if it reaches $\theta$ early. In this paper, the reliability $\gamma$ is defined as
$\gamma = \frac{1}{\sum_{i=2}^{\rm{N}}\rho_{i}}$, where 
$\rho_{i} = 1$ if $\mathrm{RMSE}_{i}\geq  \mathrm{RMSE}_{i-1}+\beta$, and $\rho_{i} = 0$ otherwise, where $\rho_{i}$ is the mark bit at the $i$th index of the UAV trajectory, $\beta$ is the acceptable RMSE gap (assumed to be 1~m in this paper), and $\mathrm{RMSE}_{i}$ is the RMSE value at the $i$th index. The reliability of the algorithms can be used as an auxiliary means to select the optimal algorithm in various environments.

\begin{figure*}[!ht]
\centering
\begin{subfigure}{0.65\columnwidth}
\centering
\includegraphics[width=\textwidth]{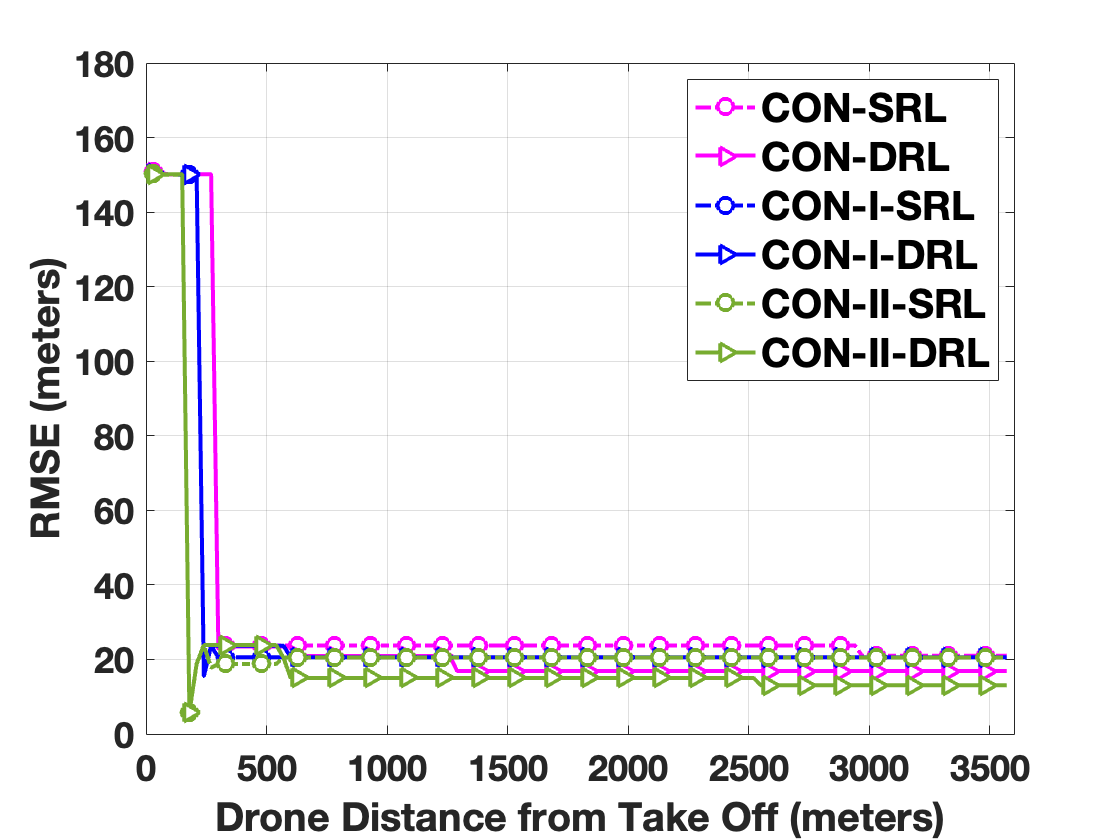}
\caption{LLS-CON, LLS-CON-I, LLS-CON-II}\label{fig:rsa}
\end{subfigure}
\begin{subfigure}{0.65\columnwidth}
\centering
\includegraphics[width=\textwidth]{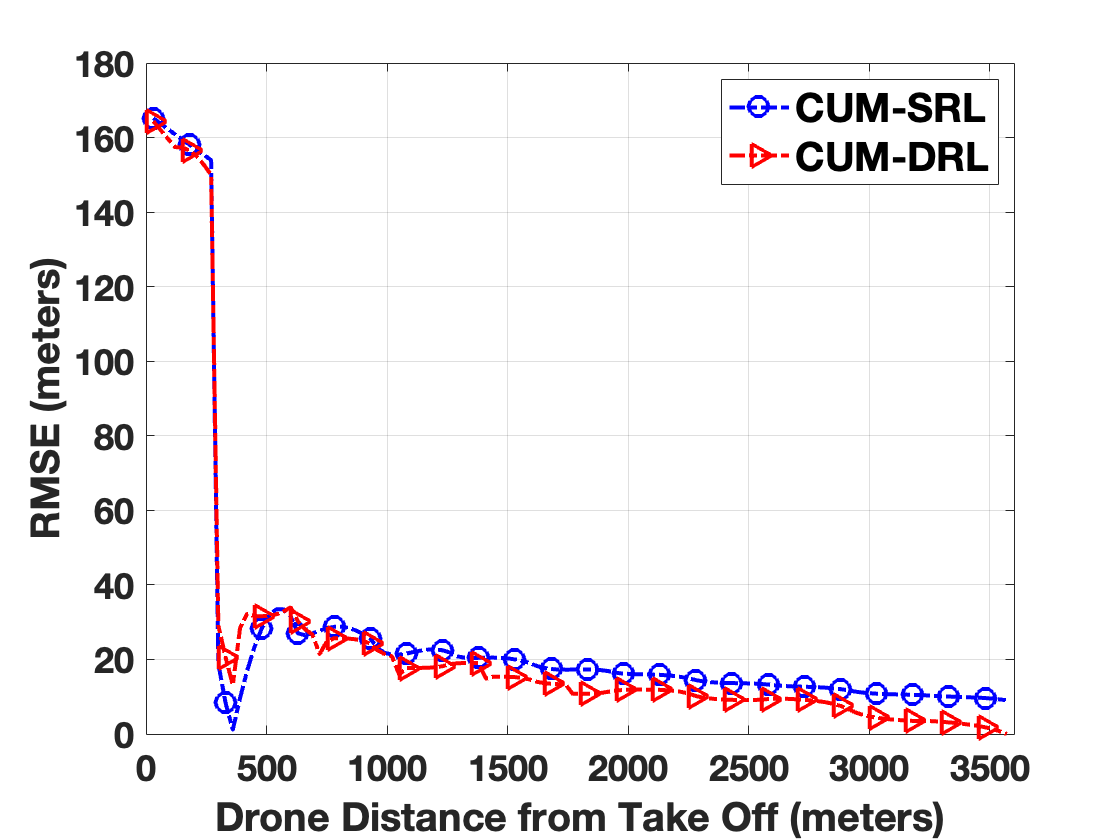}
\caption{LLS-CUM}
\label{fig:rsb}
\end{subfigure}
\begin{subfigure}{0.65\columnwidth}
\centering
\includegraphics[width=\textwidth]{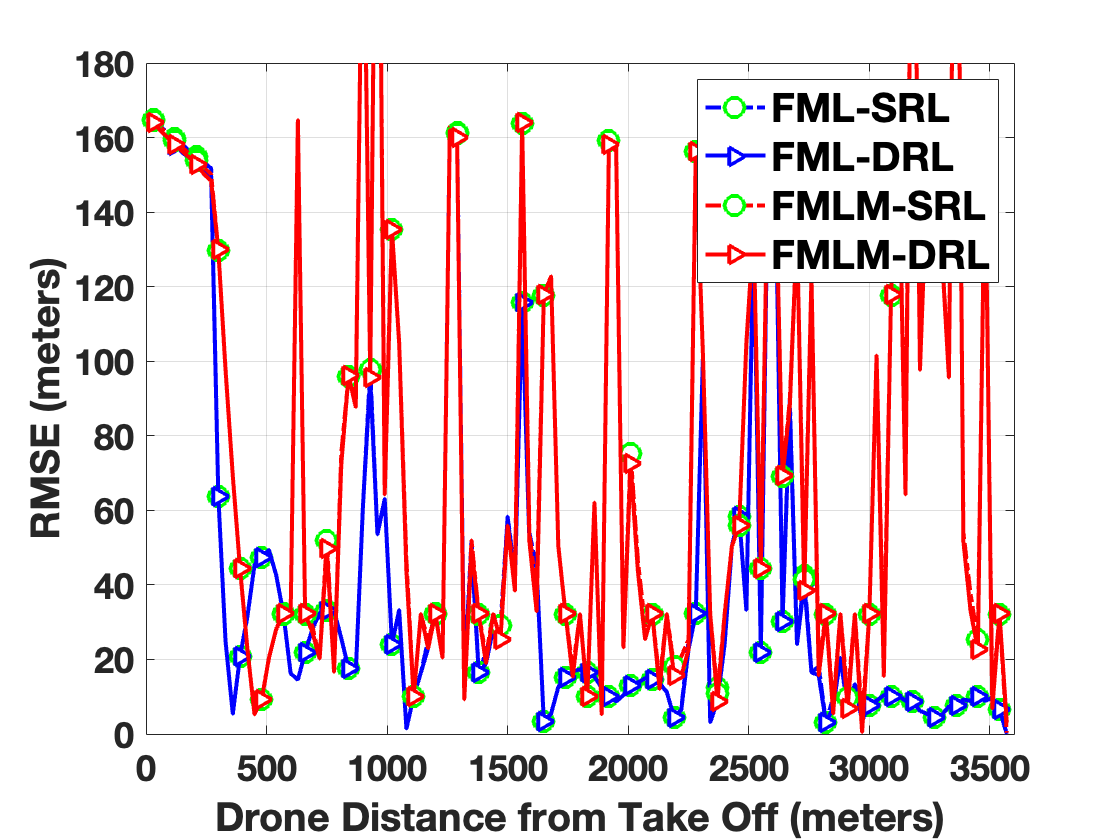}
\caption{LLS-FML and LLS-FMLM}\label{fig:rsc}
\end{subfigure}
\begin{subfigure}{0.65\columnwidth}
\centering
\includegraphics[width=\textwidth]{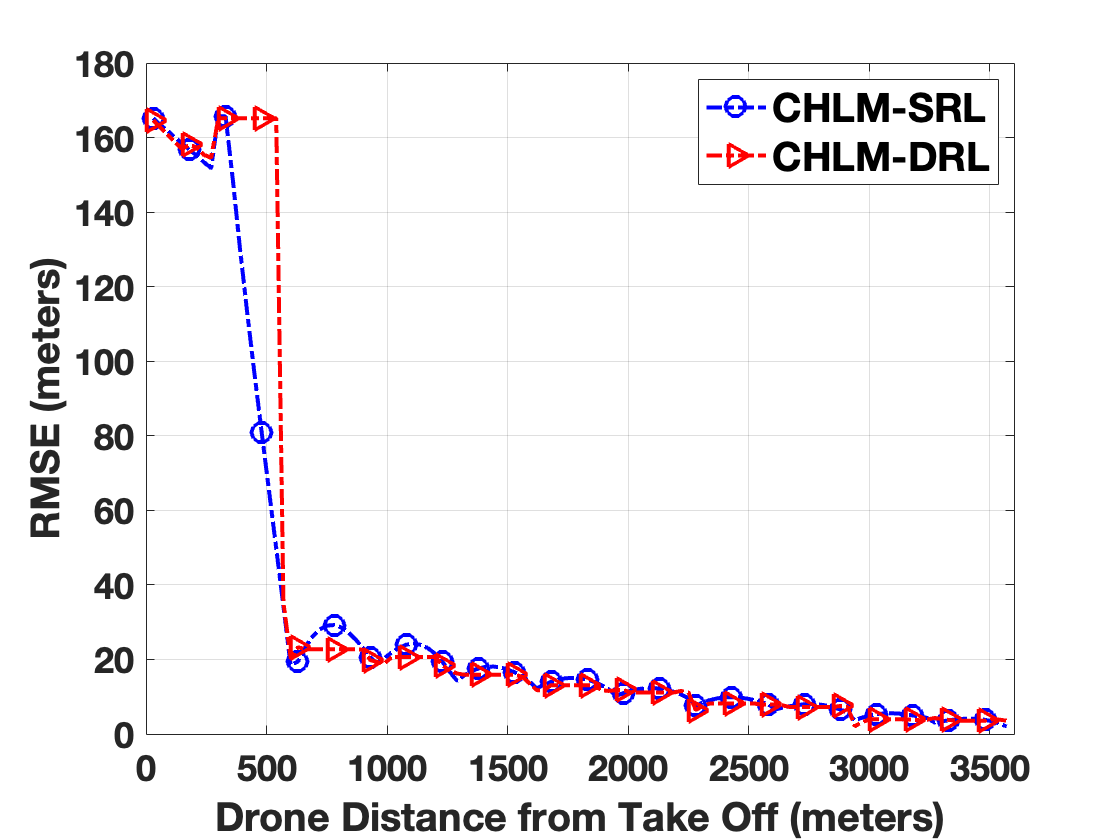}
\caption{LLS-CHLM}\label{fig:rsd}
\end{subfigure}
\begin{subfigure}{0.65\columnwidth}
\centering
\includegraphics[width=\textwidth]{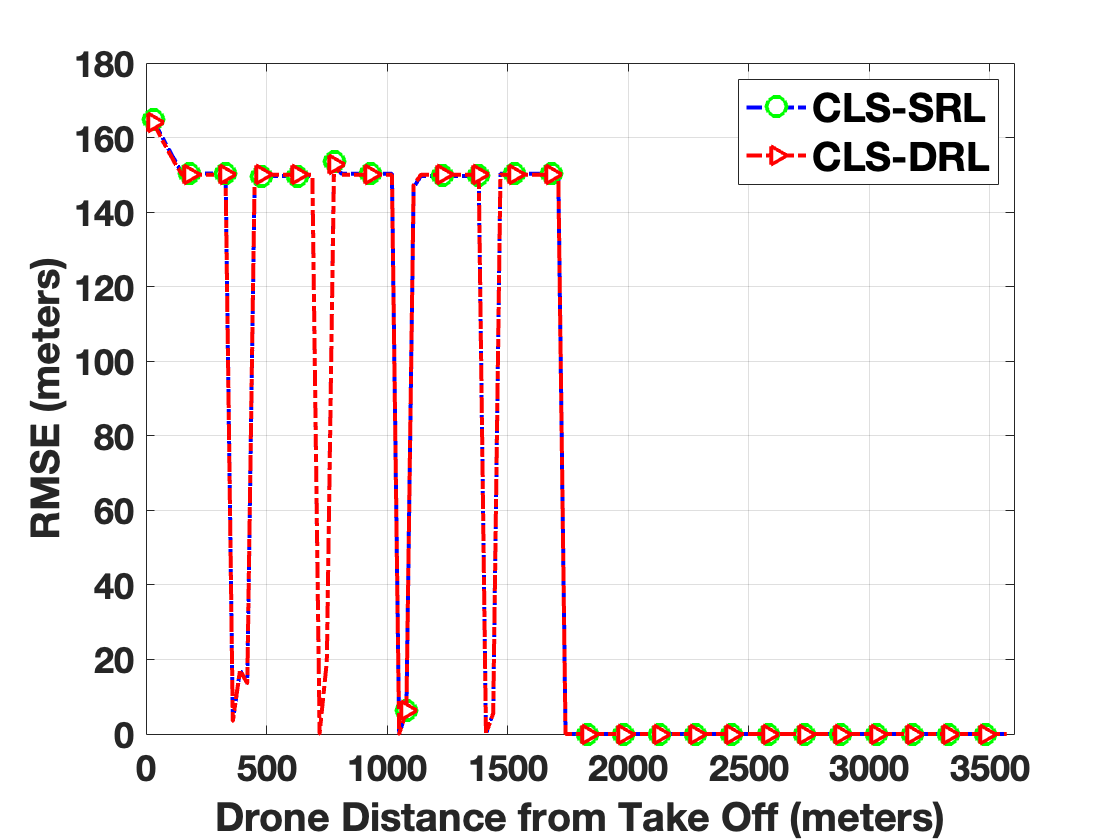}
\caption{LLS-CLS}\label{fig:rse}
\end{subfigure}
\caption{Comparison of SRL and DRL for a fixed target location considering five different SSSL approaches and their variations.}
\label{fig:rs}
\end{figure*}

\section{Numerical Results}\label{Sec:NumResult}

To evaluate the performance of the proposed approaches, a predefined trajectory of a UAV is considered. 
In particular, we consider the parallel track pattern as shown in Fig.~\ref{fig:trajec} which is one of the representative search patterns used when a uniform search is required for a large search area~\cite{drone,survel}. The UAV flies for $30~m$ between each waypoint while tracing the $300~\times~300$~m open area using the parallel track pattern. There therefore $121$~UAV waypoints (indices $i$) over the UAV's trajectory, each of which forms the candidates for the set $\boldsymbol{\upsilon}$.  
The frequency and bandwidth in this study are set to be $2.4$~GHz and $20$~MHz, respectively~\cite{uavcom}.
We use two different target location settings. First, the target location is fixed at the center of the map $\boldsymbol{T}(150,150,0)$~m, and we compare the performance difference between SRL and DRL for different approaches. Second, we consider that the target can be located anywhere on a $11\times 11$ grid of resolution on the map, and we average the performance over all the target locations. 

\begin{figure*}[!t]
\centering
\begin{subfigure}{0.65\columnwidth}
\centering
\includegraphics[width=\textwidth]{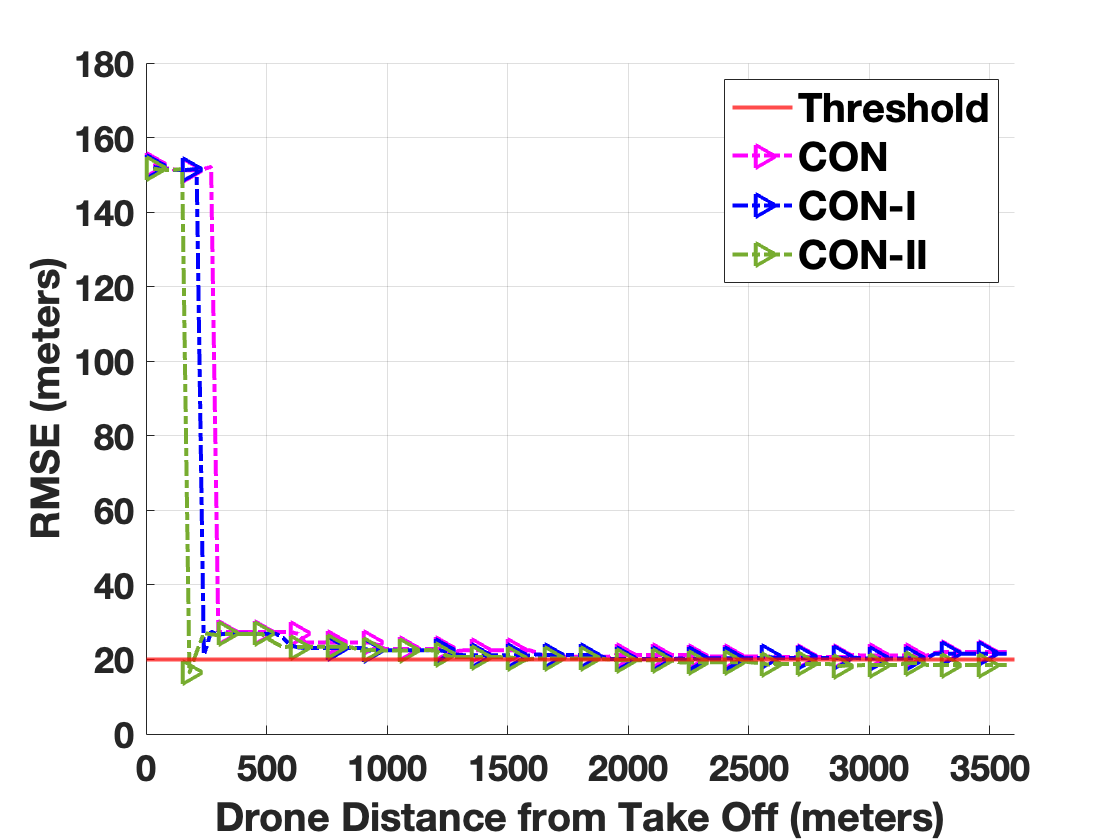}
\caption{LLS-CON, LLS-CON-I, LLS-CON-II}\label{fig:avgala}
\end{subfigure}
\begin{subfigure}{0.65\columnwidth}
\centering
\includegraphics[width=\textwidth]{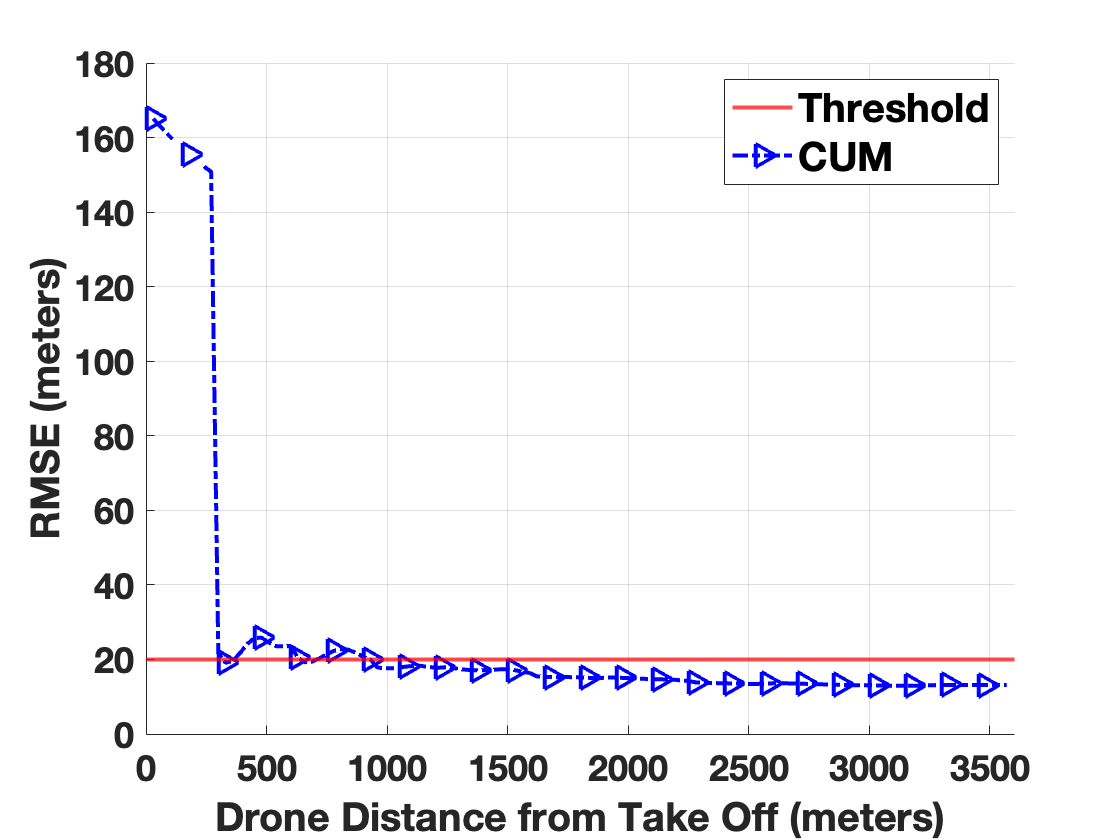}
\caption{LLS-CUM}
\label{fig:avgalb}
\end{subfigure}
\begin{subfigure}{0.65\columnwidth}
\centering
\includegraphics[width=\textwidth]{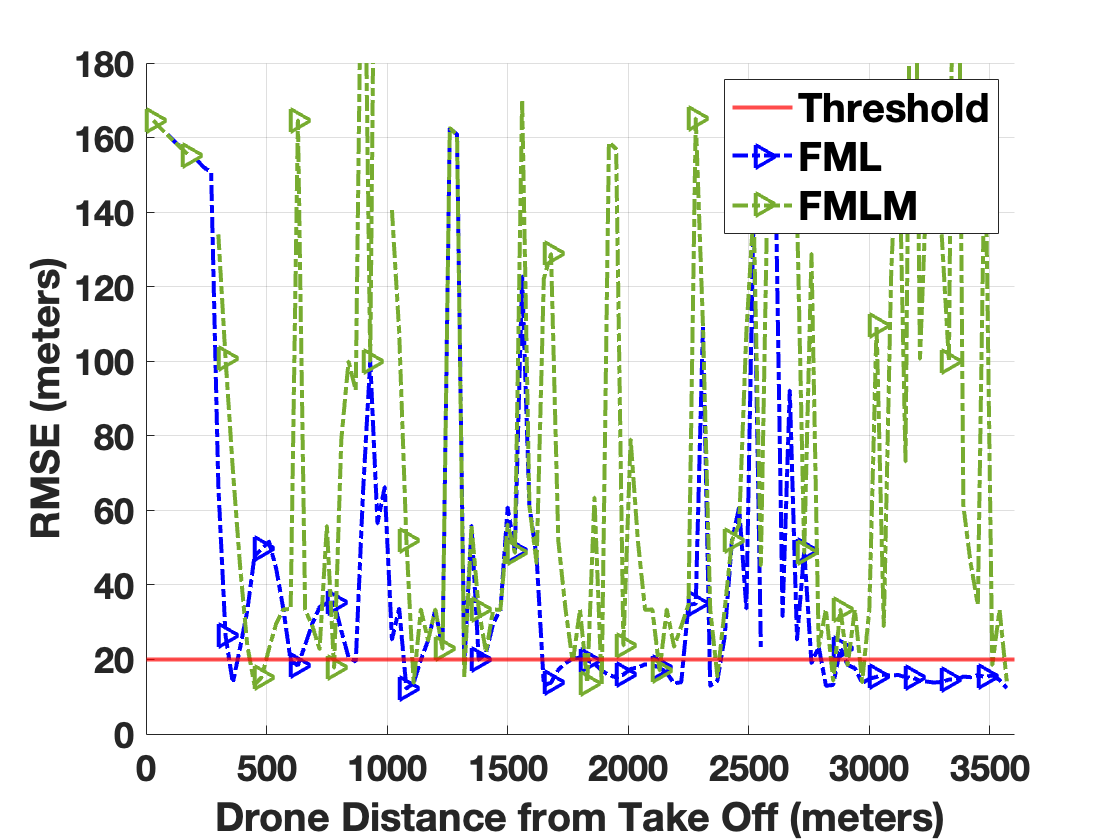}
\caption{LLS-FML and LLS-FMLM}\label{fig:avgalc}
\end{subfigure}
\begin{subfigure}{0.65\columnwidth}
\centering
\includegraphics[width=\textwidth]{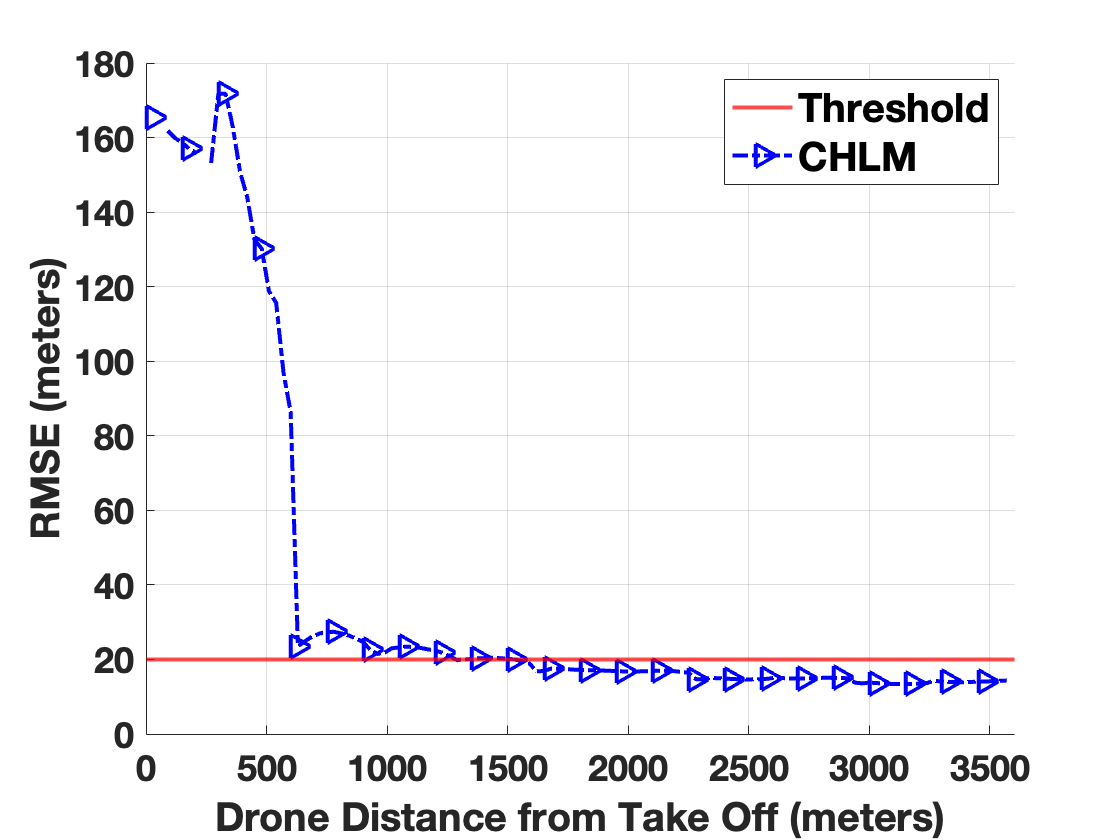}
\caption{LLS-CHLM}\label{fig:avgald}
\end{subfigure}
\begin{subfigure}{0.65\columnwidth}
\centering
\includegraphics[width=\textwidth]{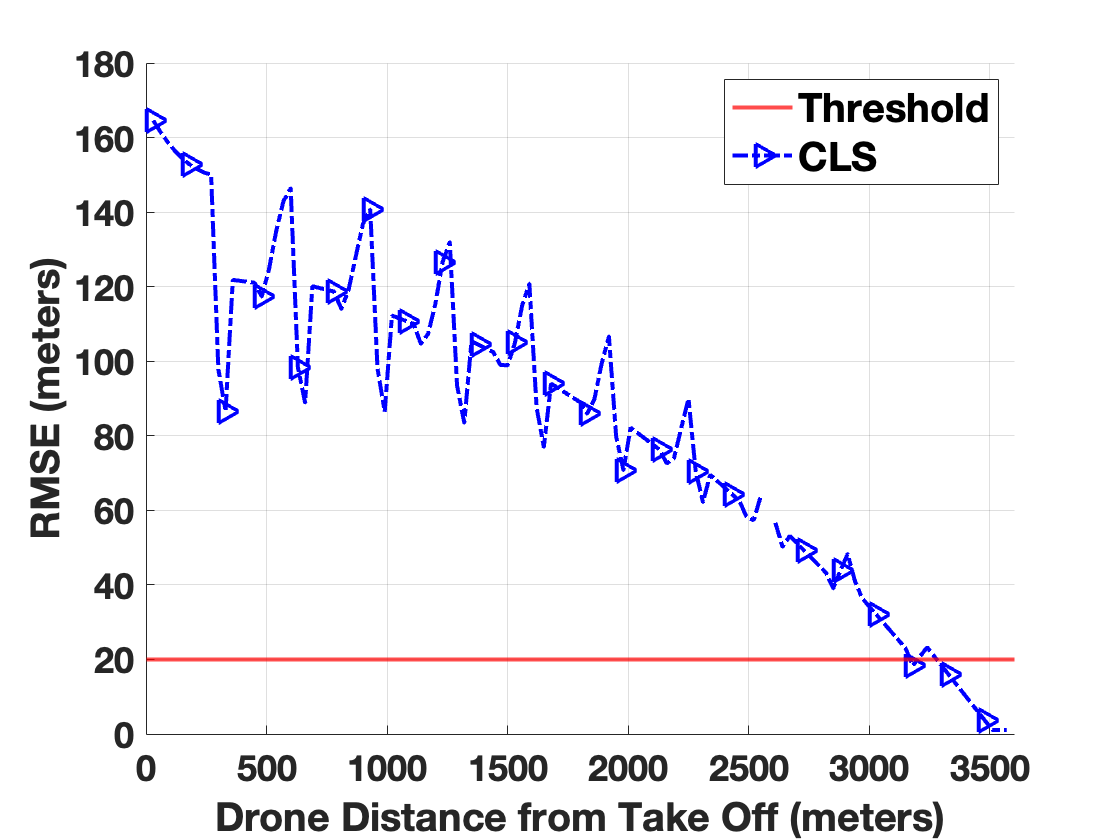}
\caption{LLS-CLS}\label{fig:avgale}
\end{subfigure}
\caption{RMSE of five different SSSL approaches and their variations averaged over different target locations.}
\label{fig:avgal}
\vspace{-0.3cm}
\end{figure*}

\begin{figure}[t!]
\centering
\includegraphics[trim=0.1cm 0.1cm 0.1cm 0.1cm, clip,width=7.5cm]{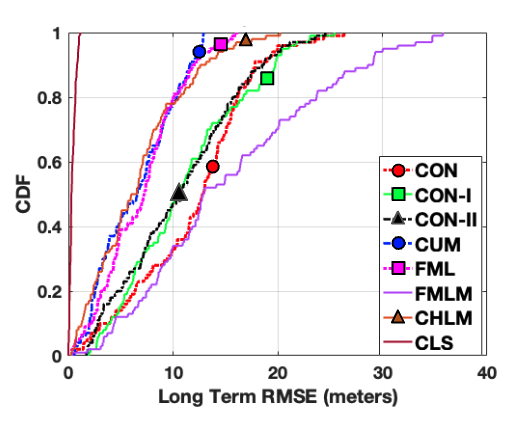}
\caption{CDF of RMSE averaged over different target locations.}
\label{fig:cdf}
\end{figure}



\subsection{Localization Performance for Fixed Target Location}

Fig.~\ref{fig:rs} demonstrates the localization performance of SRL and DRL with the five different SSSL algorithms described in Section~\ref{Sec:PropScheme}, and their variations, when the target location is fixed. The results show that LLS-CON, LLS-CUM, and LLS-CHLM show better localization performance when working with DRL. However, the difference between DRL and SRL is negligible for FDL and CLS approaches. Overall, DRL consistently provides better or similar accuracy compared to SRL for reference selection in LLS linearization. Therefore, DRL is employed for the rest of the simulations in this section. The results also show that while the accuracy tends to improve for all scenarios as the UAV moves over its trajectory, for some approaches the accuracy may get worse when $\boldsymbol{\upsilon}$ is updated based on new observations. 

\subsection{Average Localization Accuracy}
Fig.~\ref{fig:avgal} shows the localization performance of each SSSL algorithm based on their RMSE, averaged over the $11\times 11$ ground target grid. We observe that LLS-CON and LLS-CON-I cannot reach the RMSE threshold $\theta$ in Fig.~\ref{fig:avgala}. LLS-CON-II, LLS-CUM, LLS-FML, and LLS-CHLM show better accuracy with the RMSE range of 12~m to 18~m in Fig.~\ref{fig:avgala}, Fig.~\ref{fig:avgalb}, Fig.~\ref{fig:avgalc}, and Fig.~\ref{fig:avgald}, respectively. In Fig.~\ref{fig:avgale}, the LLS-CLS shows the best accuracy as the RMSE converges to 0 at the end. Additionally, Fig.~\ref{fig:cdf} shows the performance of each SSSL algorithm based on the CDF of long-term RMSE. Similar to the results of Fig.~\ref{fig:avgal}, LLS-CLS shows the best performance with the accuracy basis.

When the performance is evaluated based on the UAV flight distance, LLS-CON and LLS-CON-I cannot not reach the RMSE threshold $\theta$ even with $3,630~m$ of flight distance which is the maximum flight distance in the given map setting. LLS-CON-II, LLS-FML, and LLS-CLS which show good accuracy require at least $2,580~m$, $2,880~m$, and $3,300~m$ of flight distances to reach the desired accuracy. In contrast, LLS-CUM and LLS-CHLM require $960~m$ and $1,530~m$ flight distances and are the two quickest SSSL approaches. In terms of reliability, LLS-CON has the best reliability metric. LLS-CUM and LLS-CHLM show relatively better reliability than LLS-FML and LLS-CLS. In Table.~\ref{table1}, we summarize the performances of all the SSSL approaches considering different evaluation criteria.

\begin{table}[t]\label{t}
\caption{SSSL algorithms performance summary.}
\label{table1}
\begin{center}
\scalebox{0.68}{
\begin{tabular}{|p{0.7in}|p{0.4in}|p{0.5in}|p{1.5in}|p{0.5in}|}
\hline
\textbf{Algorithms} & $\|\boldsymbol{\upsilon}\|$ & \multicolumn{1}{c|}{\textbf{\begin{tabular}[c]{@{}c@{}}UAV flight \\ distances (m)\end{tabular}}} & \textbf{Accuracy (m) (Var (m$^2$))} & \multicolumn{1}{c|}{\textbf{Reliability}} \\ \hline
CON                 & $3$                                                                   & $-$                                                                            & 21.4659 (31.5793)                                & 1.000                    \\ \hline
CON-I                 & $3$                                                                   & $-$                                                                           & 21.5849 (33.1869)                   & 0.500                    \\ \hline
CON-II                 & $3$                                                                   & 2,580                                                                            & 17.965 (33.5379)                                & 0.500                    \\ \hline
CUM                 & $3-S$                                                                 & 960                                                                                               & 13.115 (13.7559) & 0.250                                     \\ \hline
FML                 & $3$                                                                   & 2,880                                                                                             & 12.283 (14.6101)                                & 0.025                                     \\ \hline
FMLM                 & $3-5$                                                                   & $-$                                                                                             &14.1351 (75.1869)                                & 0.021                                     \\ \hline
CHLM                & $3-5$                                                           & 1,530                                                                                             & 14.332 (20.7313)                                & 0.250                                     \\ \hline
CLS                 & $3$                                                                   & 3,300                                                                                             & 1.067 (0.0843)                                & 0.029                                     \\ \hline

\end{tabular}}
\end{center}
\vspace{-5mm}
\end{table}

\section{Conclusions}\label{Sec:Conclusion}
In this paper, we investigate the SSSL problem by considering five different approaches for selecting anchor locations on a UAV's trajectory. Using computer simulations, we show that LLS-CLS shows the best performance in terms of localization accuracy. However, it requires a relatively long flight distance, and hence, a longer search time. On the other hand,  LLS-CUM has the best performance based on the flight distance. However, it requires high computational complexity due to the increasing size of parameter matrices that are used for finding the location estimate. With this analysis, the LLS-CHLM shows to have a fairly good performance trade-off for accuracy, flight distance, and reliability, and it requires a  low computational complexity. 
Our future work includes studying these algorithms with realistic antenna characteristics and propagation conditions. We will also test the proposed approaches in a practical real-world UAV testbed, by using the emulation and testbed environments of the NSF AERPAW platform at NC State University.  

\bibliographystyle{IEEEtran}
\bibliography{references}

\vspace{12pt}

\end{document}